\documentclass[10pt,a4,leqno]{article}
\usepackage[dvips]{graphicx,psfrag}
\usepackage{rotating} 
\usepackage{amsmath}
\usepackage{amsfonts}
\usepackage{amssymb}
\usepackage{textcomp}
\usepackage{color} 
\usepackage{graphics}
\usepackage{natbib} 
\usepackage{calc,pifont,psfrag,amsthm,bbm,multirow}
\usepackage[]{caption} 

\usepackage{endnotes}

\let\footnote=\endnote

\definecolor{Gray}{rgb}{0.5,0.5,0.5}

 \usepackage{setspace}

\title{{\normalsize Updated version to appear in {\bf Mathematical Finance}}\\ --- \\ An exact formula for default swaptions' pricing in the SSRJD stochastic intensity model\thanks{We are grateful to Tomasz R. Bielecki and an anonymous referee for reading this paper and for helpful comments and suggestions.}}

\author{Damiano Brigo\thanks{Fitch Solutions and Imperial College dept. of Mathematics, 101 Finsbury Pavement, EC2A 1RS London. damiano.brigo@fitchsolutions.com} \and Naoufel
El-Bachir\thanks{ICMA Centre, The University of Reading, PO Box 242,
Reading RG6 6BA, UK. Tel:+44 (0)118 931 8239.
n.el-bachir@icmacentre.ac.uk}}

\begin{document}
\maketitle

 \newtheorem{theorem}{Theorem}[section]
 \newtheorem{lemma}{Lemma}[section]
\newtheorem{corollary}{Corollary}[section]
\newtheorem{proposition}{Proposition}[section]

\numberwithin{equation}{section}

We develop and test a fast and accurate semi-analytical formula for single-name
default swaptions in the context of a shifted square root jump
diffusion (SSRJD) default intensity model. The model can be calibrated to the CDS term structure and a few default swaptions, to price and hedge other credit derivatives consistently. We show with numerical experiments that the model implies plausible volatility smiles.


\section{Introduction}
Default swaptions are options on default swaps, hence they are often
treated by drawing analogies with interest rates swaptions,
especially as far as their Black-like market pricing formula is
concerned. Indeed, the most widely discussed model for their
valuation, the Log-normal Default Swap Market (LDSM) model is
similar to the Log-normal Swap Market (LSM) model used in interest
rates derivatives markets.

Sch\"{o}nbucher (2000) introduces the notion of survival
pricing measures by conditioning on no-default up to a given
maturity, although in contrast with the interest rate models this
measure is not equivalent to the risk neutral one. The
standard market formula for default swaptions is obtained by
modeling the default swap spread directly as a geometric brownian
motion, as summarized in Sch\"{o}nbucher (2004). The convenient
pricing measure is termed the survival swap measure whose associated ``num\'{e}raire" is the defaultable annuity which may vanish.

Jamshidian (2004) partly addresses this problem, presenting
a more formalized setup of the LDSM model, and generalizes the
theory to semi-martingales driven money market account and
conditional survival probabilities. Hull and White (2003) presents
practical aspects for the implementation of the Black market formula
and provides empirical estimates of default swap spread volatilities
for actively traded A-rated names that ranges from \%67 to \%130 in
the period from end 1999 to mid-2002. Brigo (2005) introduces
various different candidate formulations by using alternative
definitions of defaultable forward rates and develops a market model
leading to the standard Black formula under equivalent pricing
measures, showing market implied volatilities of the order of \%50 .

The variants of the Black (1976) formulae obtained for the LDSM
model, by their inherent simplicity are particularly convenient
for quoting single default swaptions by selecting an appropriate
volatility parameter. However, quoting
default swaptions for different sets of maturities and strikes or more complex
instruments consistently with just the implied volatilities given
by the formula inversion becomes problematic. One is confronted
with the need to develop a fully specified dynamical model to
impose a structure on the joint dynamics of one-period rates or
credit default swap spreads as it is done in the interest rates
derivatives markets with the LFM, or the LSM.

In particular, the
valuation of more exotic instruments like Bermudan default
swaptions requires the use of a model that accurately incorporates
the term structure of default swap rates, as well as the dynamical
deformations and movements of this term structure.  Indeed,
default swap rates are subject to large jumps and possibly
stochastic volatility effects. In the interest rates derivatives
models, these features are incorporated more or less successfully
by specifying richer joint dynamics of the forward LIBOR rates,
leading to an explosion of the number of parameters in the models.
However, contrary to the interest rate markets with their huge
number of caps/floors and swaptions, the single-name default swap
markets are most famous by the very small number of traded
instruments, rendering the calibration or estimation of any model
with a large number of parameters unfeasible.

An alternative approach, more suitable for the current state of the
default swaptions markets that is explored in this paper calls for
modeling the default intensity instead. This is the approach
followed in Brigo and Alfonsi (2003) with a stochastic default intensity model where both the short rate and the
default intensity are driven by shifted square root diffusion
processes. Brigo and Cousot (2006) examines implied volatilities
generated by this two-factor shifted square root model and
characterizes the qualitative behavior of the implied volatilities
with respect to the stochastic intensity model parameters. The
numerical experiments conducted with stylized parameter values
suggested that this model might be unable to generate large enough
implied volatilities.

Modeling the intensity process automatically
imposes a strong structure on the default swap spread joint dynamics
across different maturities, simplifying the achievement of
consistency across instruments. Judicious choices of the intensity
process can also incorporate jumps and some stochastic volatility
effects in default swap spreads, and possibly generate plausible
defaultable term structure evolutions.

In this paper, we extend the
SSRD model of Brigo and Alfonsi (2003) and Brigo and Cousot (2006)
by allowing for positive jumps in the process driving the default
intensity, consistently with empirical evidence. We develop and test a semi-analytical formula for single-name
default swaptions allowing for fast and accurate pricing.  The semi-analytical formula is based on the
celebrated decomposition due to Jamshidian (1989) for the valuation
of options on coupon bonds in one-factor affine models. We show with numerical experiments that the model generates
plausible volatility smiles. Given its relative tractability, the model can thus be calibrated to vanilla default swaptions in order to price more exotic products. A first attempt at
Bermudan default options pricing for example is in Ben Ameur et al
(2006), where the basic SSRD model is used. The jump-extended SSRJD
introduced here could be used as an improved version.

\section{The SSRJD default intensity model}
Default swap rates time series can hardly be reconciled with a
geometric brownian motion. Furthermore, forward default swaps
underlying default swaptions are not traded as such, and hence
delta-hedging can only be done approximately with spot default
swap term structures. For risk management and control purposes, it
is important to recognize the relation between different default
swaps and swaptions referencing the same credit name.

Jumps and
stochastic volatility could potentially be introduced in the
market model by postulating more appropriate dynamics
for the default swap rate. However, this would quickly destroy the
main feature of the model: its simplicity.  Also, in order to
value more exotic options, it becomes important to incorporate the
whole term structure of the default swap rates as well as
postulating dynamics that can yield appropriate deformations of
this term structure in the future.

A relatively simple candidate
for these tasks is the stochastic default intensity approach we
adopt in this paper. It models the default intensity process
instead of the default swap rate, providing a plausible approach
to consistently model the default swap rates of different
maturities. Following this approach, Brigo and Alfonsi (2003)
proposes a two-factor shifted square root diffusion model, where
both the short rate and the default intensity are assumed to
follow possibly correlated shifted square root diffusions. The
processes are modeled as a sum of a deterministic function and a
square root diffusion.

Comparing numerical examples in Brigo and
Cousot (2006) and Brigo (2005) we see that it is difficult to
produce large enough implied volatilities compared to what is
implied from default swaptions market data or historical
volatilities of default swap spreads. Hereafter we present an
extension to this model, by allowing for positive jumps in the
process driving the default intensity.

We denote the market filtration by $\mathbb{G}=(\mathcal{G}_t)_{(t
\geq 0)}$ and let $\mathbb{Q}$ be a risk-neutral probability
measure. We follow the intensity based approach to default risk
modeling and introduce the default time as a totally inaccessible
$\mathbb{G}-$stopping time $\tau.$ We further assume the usual
structure for $\mathbb{G}$, namely that $\mathbb{G}=\mathbb{F}\vee
\mathbb{H}$, where $\mathbb{F}=(\mathcal{F}_t)_{(t \geq 0)}$ is the
filtration generated by the stochastic market variables (interest
rates, default intensities, etc) except default events and
$\mathbb{H}=(\mathcal{H}_t)_{(t \geq 0)}$ is the filtration
generated by the default process: $\mathcal{H}_t =
\sigma\left(\mathbf{1}_{\{\tau < u\}}, u \leq t \right)$. It is also assumed that
there exists a strictly positive $\mathbb{F}-$adapted process
$(\lambda_t)_{(t \geq 0)}$ such that the process $(M_t)_{(t \geq
0)}$ given by:
\begin{equation}
M_t=\mathbf{1}_{\{\tau  \leq t\}}-\int_0^t \mathbf{1}_{\{\tau > s\}}
\lambda_s ds= \mathbf{1}_{\{\tau  \leq t\}}-\int_0^{t\wedge \tau }
\lambda_s ds
\end{equation}
\noindent is a uniformly integrable $\mathbb{G}-$martingale under
$\mathbb{Q}$. The process $(\lambda_t)_{(t \geq 0)}$ is referred to
as the $\mathbb{G}$ marginal intensity of the stopping time $\tau$
under $\mathbb{Q}$ or risk-neutral pre-default intensity. This setup is commonly referred to as a doubly stochastic Poisson default process or the Cox process framework. In the SSRJD
model, the intensity $\lambda_t$ is written as the sum of a positive
deterministic function $\psi(t)$ and of a positive stochastic
process $y_t$:
\begin{equation}
\lambda_t = y_t+\psi(t), \mbox{  } t \geq 0,
\end{equation}
where $\psi$ is a deterministic function of time, and is integrable
on finite intervals. The dynamics of $(y_t)_{(t \geq 0)}$ are an example of an Affine Jump Diffusion (AJD) (see Duffie et al. (2000), Duffie et al. (2003) ):
\begin{eqnarray}
dy_t &=& \kappa (\mu - y_t)dt + \nu \sqrt{y_t}dW_t + dJ_t,\\
y(0) &=& y_0, \nonumber
\end{eqnarray}
with the following condition to ensure the process cannot reach
zero:
\begin{equation}
2\kappa \mu > \nu^2.
\end{equation}
$(W_t)_{(t \geq 0)}$ is
 a Wiener
process and $(J_t)_{(t \geq 0)}$ is a pure jump process with jumps arrival rate $\alpha$ and exponentially distributed jump sizes with mean $\gamma$ preserving the attractive
feature of positive default intensity. In other terms,
$$ J_t = \sum\limits_{i=1}^{M_t} Y_i,$$ where $(M_t)$ is a Poisson
process with intensity $\alpha$, the $Y$s are independent of $M_t$ and also i.i.d. exponentially
distributed with mean $\gamma$. All the parameters $y_0,\kappa,\mu,\nu,\alpha,\gamma$ are also constrained
to positive values. Since this model belongs to the tractable AJD
class of models, the survival probability $\overline{\mathbb{S}}$
has the typical ``log-affine" shape before default:
\begin{eqnarray}
\overline{\mathbb{S}}(t,T)&=&\mathbf{1}_{\{\tau >
t\}}\mathbb{S}(t,T)=\mathbf{1}_{\{\tau >
t\}}\mathbb{E}_\mathbb{Q}\left[\exp\left(-\int_t^T \lambda_s
ds\right)|\mathcal{F}_t\right]\nonumber\\
&=& \mathbf{1}_{\{\tau >
t\}}\mathbb{E}_\mathbb{Q}\left[\exp\left(-\int_t^T[\psi(s)+y_s] ds
 \right)|\mathcal{F}_t\right]\nonumber\\
&=& \mathbf{1}_{\{\tau >
t\}}A(t,T)\exp\left(-\int_t^T\psi(s)ds-B(t,T)y_t\right),
\label{eq:survivalProb}
\end{eqnarray}
where:
\begin{eqnarray}
A(t,T) &=& \xi(t,T)\zeta(t,T),\\
B(t,T)&=&\frac{2(e^{h(T-t)}-1)}{2h+(\kappa+h)(e^{h(T-t)}-1)},
\end{eqnarray}
with $\xi(t,T)$ and $\zeta(t,T)$ given by:
\begin{eqnarray}
\xi(t,T)&=&\left(\frac{2h \exp\left(\frac{h+\kappa}{2}(T-t)\right)}{2h+(\kappa+h)(e^{h(T-t)}-1)} \right)^{\frac{2\kappa \mu}{\nu^2 }},\\
\zeta(t,T)&=&\left(\frac{2h
\exp\left(\frac{h+\kappa+2\gamma}{2}(T-t)\right)}{2h+(\kappa+h+2\gamma)(e^{h(T-t)}-1)}
\right)^{\frac{2\alpha \gamma}{\nu^2 - 2 \kappa \gamma -2 \gamma^2}},
\end{eqnarray}
and where $h=\sqrt{\kappa^2+2\nu^2}$.

Note that when $\gamma = \frac{h-\kappa}{2}$, the denominator in the exponent of $\zeta(t,T)$ goes to zero, i.e. $\nu^2 - 2 \kappa \gamma -2 \gamma^2 =0$, leading to potential numerical instabilities due to division by zero. However, one can check in this case that the base of $\zeta(t,T)$ is then equal to one. Thus for robustness of the implementation, it is necessary to set $\zeta(t,T)=1$ when $\gamma = \frac{h-\kappa}{2}$.

The SSRD model is a diffusion-only restriction of the SSRJD model
obtained by setting the jump intensity $\alpha$ to
zero, also resulting in $\zeta(t,T) =1$ in the survival probability
formula.

For default swap computations we also make use of the formula for
the following transform:
\begin{equation}
\mathbb{E}_\mathbb{Q}\left[\exp\left(-\int_t^T \lambda_s
ds\right)\lambda_T|\mathcal{F}_t\right]=-\partial_T \mathbb{S}(t,T),
\end{equation}
which can be expressed after differentiation as:
\begin{equation}
\partial_T \mathbb{S}(t,T)=
\mathbb{S}(t,T)\left[\frac{\partial_T \xi(t,T)}{\xi(t,T)}-y_t\partial_T B(t,T)+\frac{\partial_T\zeta(t,T)}{\zeta(t,T)}-\psi(T)\right],
\end{equation}
where:
\begin{eqnarray*}
\partial_T\xi(t,T)&=&\frac{-2\kappa\mu\left(e^{h(T-t)}-1 \right)}{2h+(\kappa+h)\left(e^{h(T-t)}-1\right)}\xi(t,T),\\
\partial_T B(t,T)&=&\frac{4h^2e^{h(T-t)}}{\left[2h+(\kappa+h)\left(e^{h(T-t)}-1\right)\right]^2},\\
\partial_T\zeta(t,T)&=&\frac{-2\alpha \gamma\left(e^{h(T-t)}-1
\right)}{2h+(\kappa+h+2\gamma)\left(e^{h(T-t)}-1\right)}\zeta(t,T).
\end{eqnarray*}
Again for the SSRD model, the corresponding formulae can be obtained
by simply using the fact that $\zeta(t,T)=1$ and
$\partial_T\zeta(t,T)=0$.
\section{Pricing equations for default swaps and swaptions}
\subsection{Credit Default swaps} In this section, we briefly review
default swaps pricing and refer to Brigo and Alfonsi (2003) for
further details. A (credit) default swap is a financial instrument
used by two counterparties to buy or sell protection against the
default risk of a reference credit name. In a default swap signed at
time $t$ starting at time $T_a$ with maturity $T_b$, the protection
buyer pays a periodic fee or spread $R_{a,b}(t)$ at the payment
dates $T_{a+1},\dots,T_b$ (typically quarterly) as long as the
reference entity does not default. In case of a default occurring at
time $\tau$ with $T_a < \tau \leq T_b$, the protection seller
compensates the protection buyer for his loss given default that we
assume to be a known constant $L_{GD}$. In addition, the protection
seller receives from the protection buyer the spread accrued since
the last payment date before default. In the case where $t < T_a$,
the contract is a forward default swap, while if $t=T_a$ we are
dealing with a spot default swap.

Default swaps have been shown in Brigo and Alfonsi (2003) to be
relatively insensitive to the correlation between brownians driving
the intensity and interest rate processes when both are modeled as SSRD processes, while Brigo and Cousot
(2006) confirms that it is also relatively insignificant for default
swaptions. Furthermore, Brigo and Cousot (2006) finds that the short
rate volatility has relatively little impact on the valuation of typically traded default swaptions characterized by short maturities, thus concluding that the randomness of the short rate adds
little value to stochastic intensity models for default swaptions.
Therefore, we assume a deterministic term structure of interest
rates, and denote the price at time $t$ of the default-free discount
factor for maturity $T$ or risk-free $T$-zero coupon bond by
$D(t,T)=\exp\left(-\int_t^T r_s ds\right)$.

From the perspective of a protection buyer, the value at time $t$ denoted by
$CDS(t,\Upsilon,R,L_{GD})$ of
a default swap with a payment schedule $\Upsilon=\{T_{a+1},\dots,T_b
\}$, a spread $R$ and a loss given default $L_{GD}$ is given by the following expression:
\begin{equation}
CDS(t,\Upsilon,R,L_{GD})=-\mathbf{1}_{\{\tau  > t\}}\left[R
\overline{C}_{a,b}(t)+L_{GD}\int_{T_a}^{T_b}D(t,u)\partial_u
\mathbb{S}(t,u) du\right],\label{eq:CDSValue}
\end{equation}
where \begin{equation} \overline{C}_{a,b}(t)=
\left[\sum\limits_{i=a+1}^b\alpha_i
D(t,T_i)\mathbb{S}(t,T_i)-\int_{T_a}^{T_b}(u-T_{(\beta(u)-1)})D(t,u)\partial_u
\mathbb{S}(t,u) du \right], \label{eq:RiskyBPValue}
\end{equation}
and $T_{\beta(t)}$ is the first date in the set $\{T_a,\dots,T_b \}$
that follows $t$ and $\alpha_i=T_i-T_{i-1}$ is the year fraction
between $T_{i-1}$ and $T_i$.

Hence, the fair spread $R_{a,b}(t)$ as long as default has not
occurred can be computed as the value of $R$ that equates the
default swap value to zero:
\begin{equation}
\mathbf{1}_{\{\tau  > t\}}R_{a,b}(t)=-\mathbf{1}_{\{\tau  >
t\}}\frac{L_{GD}\int_{T_a}^{T_b}D(t,u)\partial_u \mathbb{S}(t,u)
du}{\overline{C}_{a,b}(t)}.\label{eq:FairCDS}
\end{equation}

\subsection{Credit Default swaptions}
A default swaption is an option written on a default swap. In the
sequel, we will restrict the analysis to European payer default
swaptions. A payer default swaption entitles its holder the right
but not the obligation to become a protection buyer in the
underlying default swap at the expiration of the option, paying a
protection fee equal to the strike spread. Most traded single name default swaptions are canceled (or
knocked out) at default of the underlying reference name if this
occurs before the option's maturity. The
maturity of the option will typically be equal to the starting date
of the underlying default swap $T_a$. That is, the default swaption
holder enters a spot default swap if she chooses to exercise the
option at maturity.

For the pricing of a default swaption at a
valuation date $t$, the underlying is thus the $T_a$
maturity forward default swap with payment dates
$T_{a+1},\dots,T_b$. The strike $K$ specified in the contract is the
periodic fixed rate that is to be paid in exchange for the default
protection in case of exercise, instead of the
fair market spread $R_{a,b}(T_a)$ that will be available at time
$T_a$ only. The $T_a-$defaultable payoff can be valued at time $t$
by taking the risk-neutral expectation of its discounted value, where the discounting is done using the default
adjusted stochastic discount factor $\overline{D}(t,T)=\exp\left(-\int_t^T (r_s+\lambda_s) ds\right)$ as shown
in Duffie et al (1996). Hence, the payer default swaption can be
valued as in Brigo and Alfonsi (2003) and Brigo and Cousot (2006):
\begin{equation}
PSO(t,T_a,\Upsilon,K) = \mathbf{1}_{\{ \tau
> t\}}\mathbb{E}_\mathbb{Q}\left[\overline{D}(t,T_a)\overline{C}_{a,b}(T_a)\left(R_{a,b}(T_a)-K \right)^+|\mathcal{F}_t\right].
\label{eq:CDSOptValue0}
\end{equation}
A single dynamics for $R_{a,b}$ leading to a market formula
analogous to the one for interest rate swaptions is derived, under
different assumptions, in Sch\"{o}nbucher (2004) and Jamshidian
(2004). Brigo (2005) derives the same formula under different
assumptions and sketches the construction of a whole market model
for a joint family of default swap rates. Assuming that the
default swap rate $R_{a,b}$ follows a geometric brownian motion
with volatility $\sigma_{a,b}$, the above approaches allow to
price the default swaptions using Black-style formulas. Here, we
recall the formula for a payer default swaption, with self-evident
notation:
\begin{eqnarray}
PSO(t,T,\Upsilon,K)&=&\mathbf{1}_{\{ \tau >
t\}}\overline{C}_{a,b}(t)\left[R_{a,b}(t)\Phi\left(d_1\right)-K
\Phi\left(d_2\right) \right],\label{eq:market_formula}\\
d_{1,2}&=&\frac{\log{\frac{R_{a,b}(t)}{K}}\pm
(T-t)\frac{\sigma_{a,b}^2}{2}}{\sigma_{a,b}\sqrt{T-t}}.\nonumber
\end{eqnarray}
When faced with the requirement of marking a default swaption
position to market or when hedging a book of default swaptions, the need for a different model becomes apparent. Indeed the market model requires one to
input a volatility parameter. If the model could be trusted as
providing an appropriate description of the world, this parameter
(constant across maturities and strikes) could be implied from
currently traded options. Recognizing that the model is a rather
primitive approximation, one would expect to observe different
volatility parameters for different strikes and maturities
resulting in a volatility smile (or skew or smirk).

However, for a
given underlying reference name, there are often only very few
different default swaptions traded, and quite often the market is
limited to the At-The-Money (ATM) options. Deducing patterns in a
market model context can then be difficult. On the other hand, the SSRJD model can be calibrated to a default
swap rates term structure and very few default swaptions, and the
fitted values of the parameters can be used to value different
default swaptions consistently, under the condition that the model
implies meaningful patterns of implied volatilities.

To derive a semi-analytical formula for default swaptions in the SSRJD model, we use the following equivalent (to (\ref{eq:CDSOptValue0})) valuation equation:
\begin{equation}
PSO(t,T_a,\Upsilon,K)
=D(t,T_a)\mathbb{E}_\mathbb{Q}\left[\left(CDS(T_a,\Upsilon,K,L_{GD})\right)^+|\mathcal{G}_t\right].
\label{eq:CDSOptValue}
\end{equation}

Brigo and Alfonsi (2003) proposes a formula for solving this pricing
equation in the case of the SSRD model. The formula is based on the
insightful decomposition of Jamshidian (1989), where in a 1-factor
yield curve model, an option on a portfolio of cash flows is
decomposed in a portfolio of options on each cash flow, where the
strike for each option is judiciously adjusted. In the next section,
we prove and extend this formula for the SSRJD model.

\section{Analytical formula for default swaptions pricing}
The derivation of the formula follows three main steps. In
proposition \ref{prop:pricing}, we rewrite the pricing equation (\ref{eq:CDSOptValue}) in a
suitable form for the application of the decomposition, i.e. as an option on an integral of multiples of survival probabilities. Then we use
our decomposition in corollary \ref{cor:decomposition}, resulting in
the appearance of an integral of terms that are akin to options on
survival probabilities. And lastly, we give an explicit formula for
these options in proposition \ref{prop:formulas}. Note that proposition \ref{prop:pricing} and corollary \ref{cor:decomposition} are model-independent in the following sense. Proposition \ref{prop:pricing} holds for any nonnegative default intensity process $(\lambda_t)$ and corollary \ref{cor:decomposition} requires the additional assumption of a survival probability function that is monotonic in $\lambda_0$. We only really use the SSRJD dynamics to derive the formula in proposition \ref{prop:formulas}.
\begin{proposition}
The default swaption price satisfies the following
formula:
\begin{equation}PSO(t,T_a,\Upsilon,K) = \mathbf{1}_{\{ \tau
> t\}}D(t,T_a)\mathbb{E}_\mathbb{Q}\left[e^{-\int_t^{T_a}\lambda_s ds}\left(L_{GD}-\int_{T_a}^{T_b} h(u)\mathbb{S}(T_a,u)du \right)^+|\mathcal{F}_t\right],
\label{eq:pso}
\end{equation}
\normalsize where $h$ is defined as:
\begin{equation}
h(u) = D(T_a,u)\left[L_{GD}\bigl(r_u+\delta_{T_b}(u)\bigr)+
K\bigl(1-(u-T_{\beta(u)-1})r_u\bigr) \right],
\end{equation}
with $\delta_{T_b}(u)$ the Dirac delta function centered at $T_b$.
\label{prop:pricing}
\end{proposition}

\begin{proof}
Starting from equation (\ref{eq:CDSOptValue}), we substitute the
default swap $T_a-$value from equations (\ref{eq:CDSValue}) and
(\ref{eq:RiskyBPValue}) resulting in the following formula:\small
\begin{eqnarray*}
PSO(t,T_a,\Upsilon,K)&=&D(t,T_a)\mathbb{E}_\mathbb{Q}\Biggl[\mathbf{1}_{\{
\tau
>
T_a\}}\Biggl(K\int_{T_a}^{T_b}D(T_a,u)(u-T_{\beta(u)-1})\partial_u \mathbb{S}(T_a,u)du\\
&-&K\sum\limits_{i=a+1}^b\alpha_i
D(T_a,T_i)\mathbb{S}(T_a,T_i)-L_{GD}\int_{T_a}^{T_b}D(T_a,u)\partial_u
\mathbb{S}(T_a,u)du\Biggr)^+ |\mathcal{G}_t\Biggr].
\end{eqnarray*}\normalsize
We can integrate by parts the last integral of the above expression:
\begin{eqnarray*}
\int_{T_a}^{T_b}D(T_a,u)\partial_u \mathbb{S}(T_a,u)du &=&
\Bigl[D(T_a,u)\mathbb{S}(T_a,u)
\Bigr]_{T_a}^{T_b}-\int_{T_a}^{T_b}\mathbb{S}(T_a,u)\partial_u
D(T_a,u)
du\\
&=&D(T_a,T_b)\mathbb{S}(T_a,T_b)-1-\int_{T_a}^{T_b}\mathbb{S}(T_a,u)\partial_u
D(T_a,u) du.
\end{eqnarray*}
For the other integral appearing in the default swaption price, we
first decompose it in a sum of integrals where the limits of
integration are the default swap payment dates:
\begin{equation*}
\int_{T_a}^{T_b}D(T_a,u)(u-T_{\beta(u)-1})\partial_u
\mathbb{S}(T_a,u)du
=\sum\limits_{i=a}^{b-1}\int_{T_i}^{T_{i+1}}D(T_a,u)(u-T_i)\partial_u
\mathbb{S}(T_a,u)du,
\end{equation*}
where we used the fact that for $T_i < u < T_{i+1}$, $T_{\beta(u)-1}
= T_i$. And we can now integrate by parts these integrals:
\begin{eqnarray*}
\int_{T_i}^{T_{i+1}}D(T_a,u)(u-T_i)\partial_u \mathbb{S}(T_a,u)du&=&\Bigl[D(T_a,u)(u-T_i)\mathbb{S}(T_a,u)\Bigr]_{T_i}^{T_{i+1}}\\
&-&\int_{T_i}^{T_{i+1}}D(T_a,u)\mathbb{S}(T_a,u)du\\
&-&\int_{T_i}^{T_{i+1}}\mathbb{S}(T_a,u)(u-T_i)\partial_u D(T_a,u)du.
\end{eqnarray*}
Using the fact that $T_{i}-T_{i-1}=\alpha_i$, we obtain after
summation:
\begin{eqnarray*}
\int_{T_a}^{T_b}D(T_a,u)(u-T_{\beta(u)-1})\partial_u
\mathbb{S}(T_a,u)du
&=& \sum\limits_{i=a+1}^{b}\alpha_i D(T_a,T_i)\mathbb{S}(T_a,T_i)\\
&-&\int_{T_a}^{T_{b}}D(T_a,u)\mathbb{S}(T_a,u)du\\
&-&\int_{T_a}^{T_{b}}\mathbb{S}(T_a,u)(u-T_{\beta(u)-1})\partial_u
D(T_a,u)du.
\end{eqnarray*}
Note that $\partial_u D(T_a,u)=-r_u D(T_a,u)$, substitute the
expressions obtained for the integrals back in the original formula,
using
$$D(T_a,T_b)\mathbb{S}(T_a,T_b)=\int_{T_a}^{T_b}D(T_a,u)\mathbb{S}(T_a,u)\delta_{T_b}(u)du,$$
and finally, using the formula
$$\mathbb{E}_\mathbb{Q}[\mathbf{1}_{\{ \tau
>
T_a\}} Y_{T_a}|\mathcal{G}_t]= \mathbf{1}_{\{ \tau
>
t\}}\mathbb{E}_\mathbb{Q}\left[\exp\left(-\int_t^{T_a}\lambda_s
ds\right) Y_{T_a}|\mathcal{F}_t\right]$$ (see for example Bielecki
and Rutkowski (2001), Corollary 5.1.1 p.145), we obtain the result of
the proposition after rearranging.
\end{proof}
Jamshidian (1989) decomposes an option on a portfolio of zero-coupon
bonds in a portfolio of options on the zero-coupon bonds. The
rewriting of the pricing problem as in equation (\ref{eq:pso}) will
now allow us to achieve a similar result. Indeed, the term
$\int_{T_a}^{T_b} h(u)\mathbb{S}(T_a,u) du$ is akin to a portfolio
of survival probabilities of infinitely many maturities. We also
note that survival probabilities satisfy the same formulas as
zero-coupon bonds where the default intensity plays the role of the
short rate. Hence, the expectation in equation (\ref{eq:pso}) can be
seen as a put option on a portfolio of zero-coupon bonds (although
with infinitely many) where the strike is $L_{GD}$ and the interest
rate is given by the default intensity $\lambda_t$. Therefore, it is
only natural that we are able to decompose it as a portfolio of
infinitely many options on survival probabilities.
\begin{corollary}
Assume the short rate $r_u$ is nonnegative and bounded by $100\%$: $0 \leq r_u \leq 1 \mbox{ }\forall u \in [0,T_b]$ and that the spread payments occur at least once a year\footnote{Usually spreads are paid quarterly.} such that $0 \leq u-T_{\beta(u)-1} \leq 1$.
If the following integral is positive
\begin{equation}
\int_{T_a}^{T_b}\Biggl[ L_{GD} D(T_a,u)
\partial_u \mathbb{S}(T_a,u;0)
+K\mathbb{S}(T_a,u;0)D(T_a,u)\left(1-(u-T_{\beta(u)-1})r_u
\right)\Biggr]du,
\end{equation}
then the default swaption price is
the solution to the following
formula:
\begin{equation}\mathbf{1}_{\{ \tau
>
t\}}D(t,T_a)\int_{T_a}^{T_b}h(u)\mathbb{E}\left[\exp\left(-\int_t^{T_a}\lambda_s
ds\right)\bigl(\mathbb{S}(T_a,u;y^*)-\mathbb{S}(T_a,u;y_{T_a})\bigr)^+|\mathcal{F}_t\right]du,
\label{eq:pso2}
\end{equation}
where $y^* \geq 0$ satisfies:
\begin{equation}
\int_{T_a}^{T_b} h(u)\mathbb{S}(T_a,u;y^*) du = L_{GD}.
\label{eq:Condition}
\end{equation}
Otherwise, the default swaption price is simply given by the
corresponding forward default swap value:
$$CDS(t,\Upsilon,K,L_{GD}).$$
\label{cor:decomposition}
\end{corollary}

\begin{proof}
Recall the definition of $h(u)$:
$$ h(u) = D(T_a,u)\left[L_{GD}\bigl(r_u+\delta_{T_b}(u)\bigr)+ K\bigl(1-(u-T_{\beta(u)-1})r_u\bigr)
\right].$$ Since $0 \leq r_u \leq 1$ and $0 \leq u-T_{\beta(u)-1} \leq 1$, it follows that
$$ h(u) \geq 0,  \mbox{  for all u}.$$
Also note that $h(u)$ is a deterministic function that does not
depend on $y$, while the survival probability $\mathbb{S}(T_a,u;y)$
given by equation (\ref{eq:survivalProb}) is clearly monotonically
decreasing in $y$ for all $T_a$ and $u$. Hence,$$\int_{T_a}^{T_b}
h(u)\mathbb{S}(T_a,u;y) du $$ is a monotonically decreasing function
of $y$. Furthermore, it is easy to see from equation
(\ref{eq:survivalProb}) that
$$ \lim\limits_{y\rightarrow \infty} \int_{T_a}^{T_b}
h(u)\mathbb{S}(T_a,u;y) du = 0 < L_{GD},$$ or just recall that
$\mathbb{S}(T_a,u;y)$ is a survival probability and $y$ is the
initial value of the stochastic process driving the default
intensity.

We are interested in finding if there exists $y^* \geq 0$ satisfying
equation (\ref{eq:Condition}). Now, recall that
$$h(u)= L_{GD}r_u D(T_a,u)-K(u-T_{\beta(u)-1})r_u
D(T_a,u)+L_{GD}\delta_{T_b}(u)D(T_a,u)+KD(T_a,u)$$ and note that
(integrating by parts):
$$\int_{T_a}^{T_b}r_u D(T_a,u)\mathbb{S}(T_a,u)du =
1-D(T_a,T_b)\mathbb{S}(T_a,T_b)+\int_{T_a}^{T_b} D(T_a,u)\partial_u
\mathbb{S}(T_a,u)du.$$

Hence, substituting back in the original integral, we obtain the
following: \begin{eqnarray*} \int_{T_a}^{T_b} h(u)\mathbb{S}(T_a,u)
du &=& L_{GD}+\int_{T_a}^{T_b}\Biggl[ L_{GD} D(T_a,u)
\partial_u \mathbb{S}(T_a,u)\\
&+&K\mathbb{S}(T_a,u)D(T_a,u)\left(1-(u-T_{\beta(u)-1})r_u
\right)\Biggr]du,
\end{eqnarray*}
so that: \begin{eqnarray*} \lim\limits_{y\rightarrow 0^+}
\int_{T_a}^{T_b} h(u)\mathbb{S}(T_a,u;y) du &=& L_{GD} +
\int_{T_a}^{T_b}\Biggl[ L_{GD} D(T_a,u)
\partial_u \mathbb{S}(T_a,u;0)\\
&+&K\mathbb{S}(T_a,u;0)D(T_a,u)\left(1-(u-T_{\beta(u)-1})r_u
\right)\Biggr]du.
\end{eqnarray*}

Note that $\left(1-(u-T_{\beta(u)-1})r_u \right)$
is nonnegative since $r_u\leq1$ and $(u-T_{\beta(u)-1})\leq 1$. And
since ${S}(T_a,u;0)$ and $D(T_a,u)$ are both nonnegative being
respectively the survival probability and the discount factor at
time $T_a$ for maturity $u$, it follows that:
$$K\mathbb{S}(T_a,u;0)D(T_a,u)\left(1-(u-T_{\beta(u)-1})r_u
\right) \geq 0 \mbox{  for all u } \geq T_a.$$ On the other hand,
$\partial_u \mathbb{S}(T_a,u;0) \leq 0$ since the survival
probability is decreasing with maturity.

We can then consider two cases depending on whether the integral
$$\int_{T_a}^{T_b}\Biggl[ L_{GD} D(T_a,u)
\partial_u \mathbb{S}(T_a,u;0)
+K\mathbb{S}(T_a,u;0)D(T_a,u)\left(1-(u-T_{\beta(u)-1})r_u
\right)\Biggr]du$$ is negative or not.

In the first case, i.e. when the integral is negative:
$$\lim\limits_{y\rightarrow 0^+}
\int_{T_a}^{T_b} h(u)\mathbb{S}(T_a,u;y) du < L_{GD},$$ and then the
equation (\ref{eq:Condition}) does not admit a solution in $y$.
However, in this case the payoff of the option is $\mathbb{Q}-a.s.$
strictly positive and hence the payoff of the option simplifies to a
forward default swap payoff.

In the other case (i.e. when the integral is nonnegative), by the
intermediate value theorem the equation (\ref{eq:Condition}) admits
a unique solution $y^*$ by continuity and monotonicity, and we can
replace $L_{GD}$ by $\int_{T_a}^{T_b} h(u)\mathbb{S}(T_a,u;y^*) du$
in (\ref{eq:pso}). Since $\mathbb{S}(T_a,u;y)$ is a monotonic
function in $y$, then the terms
$\mathbb{S}(T_a,u;y^*)-\mathbb{S}(T_a,u;y_{T_a})$ will be all of the
same sign for all values of $u$, and therefore:
\begin{equation*}
\left(\int_{T_a}^{T_b}h(u)\bigl(\mathbb{S}(T_a,u;y^*)-\mathbb{S}(T_a,u;y_{T_a})\bigr)
du\right)^+=\int_{T_a}^{T_b}h(u)\bigl(\mathbb{S}(T_a,u;y^*)-\mathbb{S}(T_a,u;y_{T_a})\bigr)^+
du,
\end{equation*}
which we can substitute back in the expression (\ref{eq:pso}) for
the default swaption value, and use Fubini's theorem to change the
order of the integrations, resulting in equation (\ref{eq:pso2}),
thus completing the proof.
\end{proof}

Having decomposed the default swaption price in terms of options on
survival probabilities, we are left with the task of computing these
option values. Indeed, to further compute the quantity given in
equation (\ref{eq:pso2}), recall that:
\begin{equation*}
\int_t^T \lambda_s ds = \int_t^T \psi(s) ds + \int_t^T y_s ds,
\end{equation*}
and that the survival probabilities $\mathbb{S}(t,T)$ satisfy
equation (\ref{eq:survivalProb}). Substituting these in formula
(\ref{eq:pso2}) results in the following expression for the default
swaption:
\begin{eqnarray}
&\mathbf{1}_{\{ \tau> t\}}D(t,T_a)\exp\left(-\int_t^{T_a}
\psi(s)ds\right)*\\
&\int_{T_a}^{T_b} h(u)A(T_a,u)e^{-\int_{T_a}^u \psi(s) ds}
\mathbb{E}\left[\exp\left(-\int_t^{T_a} y_s
ds\right)\left(e^{-B(T_a,u)y^*}-e^{-B(T_a,u)y_{T_a}}\right)^+
|\mathcal{F}_t\right]du.\nonumber
\end{eqnarray}
The above expression is analytic up to an integral if we are able to
find a formula for the expectation involved. We take up that task in
the next proposition where:
$$ \Psi(t,T,y_t,\varsigma,\varrho):= \mathbb{E}\left[\exp\left(-\int_t^{T} y_s
ds\right)\left(e^{-\varrho \varsigma}-e^{-\varrho
y_{T}}\right)^+|\mathcal{F}_t\right],$$ with nonnegative $\varsigma$ and
$\varrho$.

\begin{proposition}
\begin{equation}
\Psi(t,T,y_t,\varsigma,\varrho)=e^{-\varrho
\varsigma}\Pi(T-t,y_t,\varsigma,0)-\Pi(T-t,y_t,\varsigma,\varrho),
\end{equation}
where
\begin{equation}
\Pi(T,y_0,\varsigma,\varrho)=\frac{1}{2}\alpha_{\psi}(T)e^{-\beta_{\psi}(T)y_0}-\frac{1}{\pi}\int_0^\infty
\frac{e^{U y_0}[S \cos(W y_0+v \varsigma)+R \sin(W y_0+v
\varsigma)]}{v}dv,
\end{equation}
with
\begin{eqnarray}
\beta_{\psi}(T) &=& \frac{2\varrho
h+\left(2+\varrho(h-\kappa)\right)(e^{h
T}-1)}{2h+(h+\kappa+\varrho \nu^2)(e^{h T}-1)},\label{eq:beta}\\
\alpha_{\psi}(T) &=&
\left[\frac{2h\exp\left(\frac{\kappa+h}{2}T\right)}{2h+(h+\kappa+\varrho
\nu^2)(e^{h
T}-1)} \right]^{\frac{2\kappa \mu}{\nu^2}}\nonumber\\
&*&\left[\frac{2h(1+\varrho
\gamma)\exp\left(\frac{(h^2-(\kappa+2\gamma)^2)(1-\frac{\varrho}{2}(h+\kappa))}{2(h-\kappa-2\gamma+\varrho(\gamma(h+\kappa)-\nu^2))}T\right)}{2h(1+\varrho
\gamma)+\left[h+\kappa+\varrho
\nu^2+\gamma(2+\varrho(h-\kappa))\right](e^{h
T}-1)}\right]^{\frac{2\alpha
\gamma}{\nu^2-2\kappa\gamma-2\gamma^2}},\label{eq:alpha}
\end{eqnarray}
and\small
\begin{eqnarray*}
R &=& (J^2+K^2)^{\frac{D}{2}}e^G[E \cos(H+D\arctan(K/J))-F
\sin(H+D\arctan(K/J))],\\
S &=& (J^2+K^2)^{\frac{D}{2}}e^G[F \cos(H+D\arctan(K/J))+E
\sin(H+D\arctan(K/J))],\\
U &=& \frac{\delta+\varepsilon e^{hT}+\phi e^{2hT}}{N},\\
W &=& -\frac{4vh^2 e^{hT}}{N},\\
\nonumber\\
E &=&(\widetilde{x}^2+\widetilde{y}^2)^{\frac{\kappa
\mu}{\nu^2}}\cos\left(\frac{2\kappa
\mu}{\nu^2}\arctan\left(\frac{\widetilde{y}}{\widetilde{x}}\right)\right),\\
F &=&(\widetilde{x}^2+\widetilde{y}^2)^{\frac{\kappa
\mu}{\nu^2}}\sin\left(\frac{2\kappa
\mu}{\nu^2}\arctan\left(\frac{\widetilde{y}}{\widetilde{x}}\right)\right),\\
\nonumber\\
\widetilde{x} &=& \frac{2h
e^{(h+\kappa)\frac{T}{2}}[2h+(h+\kappa+\varrho\nu^2)(e^{hT}-1)]}{N},\\
\widetilde{y} &=&- \frac{2h
e^{(h+\kappa)\frac{T}{2}}v\nu^2[e^{hT}-1]}{N},\\
\nonumber\\
D &=& \frac{-2\gamma\alpha}{\nu^2-2\gamma\kappa-2\gamma^2},\\
G &=& \frac{\alpha\gamma
T[(2-\varrho(h+\kappa))(h-\kappa-2\gamma-\varrho[\nu^2-\gamma(h+\kappa)])+v^2(h+\kappa)[\nu^2-\gamma(h+\kappa)]]}{(h-\kappa-2\gamma-\varrho[\nu^2-\gamma(h+\kappa)])^2+v^2[\nu^2-\gamma(h+\kappa)]^2},\\
H &=& \frac{\alpha\gamma T
v[(2-\varrho(h+\kappa))[\nu^2-\gamma(h+\kappa)]-(h+\kappa)(h-\kappa-2\gamma-\varrho[\nu^2-\gamma(h+\kappa)])
]}{(h-\kappa-2\gamma-\varrho[\nu^2-\gamma(h+\kappa)])^2+v^2[\nu^2-\gamma(h+\kappa)]^2},\\
J &=& 1 +
\frac{(e^{hT}-1)[(h+\kappa+2\gamma)(1+\varrho\gamma)+(\nu^2+\gamma(h-\kappa))[\varrho(\varrho\gamma+1)+v^2\gamma]]}{2h(1+\varrho
\gamma)^2 + 2hv^2\gamma^2},\\
K &=& -
\frac{(e^{hT}-1)v[2\gamma\kappa+2\gamma^2-\nu^2]}{2h(1+\varrho
\gamma)^2 + 2hv^2\gamma^2},\\
N &=&
(2h+(h+\kappa+\varrho\nu^2)[e^{hT}-1])^2+v^2\nu^4[e^{hT}-1]^2,\\
\delta &=&
2(h-\kappa)-4\nu^2\varrho+\varrho^2\nu^2(h+\kappa)+v^2\nu^2(h+\kappa),\\
\varepsilon &=&
4\kappa-4\kappa^2\varrho-2\kappa\varrho^2\nu^2-2v^2\nu^2\kappa,\\
\phi &=&
-2(h+\kappa)-4\nu^2\varrho-\varrho^2\nu^2(h-\kappa)-v^2\nu^2(h-\kappa).
\end{eqnarray*}\normalsize
\label{prop:formulas}
\end{proposition}

\begin{proof}
First note the equivalence between the following events:
\begin{equation*}
\{e^{-\varrho \varsigma} \geq e^{-\varrho y_{T}}  \} \Leftrightarrow
\{ y_{T} \geq \varsigma     \}.
\end{equation*}
Hence:
\begin{eqnarray*} \Psi(t,T,y_t,\varsigma,\varrho)&=&e^{-\varrho \varsigma}\mathbb{E}\left[\exp\left(-\int_t^{T} y_s
ds\right)\mathbf{1}_{\{ y_{T} \geq \varsigma\}}
|\mathcal{F}_t\right]\\
&-& \mathbb{E}\left[\exp\left(-\varrho y_T-\int_t^{T} y_s
ds\right)\mathbf{1}_{\{ y_{T} \geq \varsigma \}}
|\mathcal{F}_t\right].
\end{eqnarray*}
We define $\Pi$ as follows:
\begin{equation*}
\Pi(T,y_0,\varsigma,\varrho) :=\mathbb{E}\left[\exp\left(-\varrho
y_T-\int_0^{T} y_s ds\right)\mathbf{1}_{\{ y_T \geq \varsigma\}}
\right].
\end{equation*}
Christensen (2002) derived a formula for $\Pi$ that is analytic up
to an integral. His formula is also reported in Lando (2004) Appendix E. We recall it below:
$$\Pi(T,y_0,\varsigma,\varrho)=\frac{1}{2}\psi(T,y_0,\varrho)-\frac{1}{\pi}\int_0^\infty \frac{\mbox{Im}\left[e^{i v \varsigma}\psi(T,y_0,-\varrho-i v)\right]}{v}dv, $$
with
\begin{equation*}
\psi(T,y_0,\varrho)=\alpha_{\psi}(T)e^{-\beta_{\psi}(T)y_0},
\end{equation*}
where $\alpha_{\psi}$ and $\beta_{\psi}$ satisfy formulae
(\ref{eq:alpha}) and (\ref{eq:beta}) respectively. The imaginary
part appearing above admits an explicit expression as given in the
statement of the proposition.

Since the process $y_t$ is a homogenous and markovian jump-diffusion
\begin{eqnarray*}
\mathbb{E}\left[\exp\left(-\varrho y_T-\int_t^{T} y_s
ds\right)\mathbf{1}_{\{ y_{T} \geq \varsigma \}}
|\mathcal{F}_t\right]&=&\mathbb{E}_{y_t}\left[\exp\left(-\varrho
y_{T-t}-\int_0^{T-t} y_s ds\right)\mathbf{1}_{\{ y_{T-t} \geq
\varsigma \}}
\right]\\
&=&\Pi(T-t,y_t,\varsigma,\varrho).
\end{eqnarray*}
\end{proof}

In summary, if
\begin{equation*}
\int_{T_a}^{T_b}\Biggl[ L_{GD} D(T_a,u)
\partial_u \mathbb{S}(T_a,u;0)
+K\mathbb{S}(T_a,u;0)D(T_a,u)\left(1-(u-T_{\beta(u)-1})r_u
\right)\Biggr]du > 0,
\end{equation*}
then it is possible to solve for a positive $y^*$ satisfying
$\int_{T_a}^{T_b} h(u)\mathbb{S}(T_a,u;y^*) du = L_{GD} $, and such
that the default swaption price is given by:
\begin{equation*}
\mathbf{1}_{\{ \tau> t\}}D(t,T_a) e^{-\int_t^{T_a}
\psi(s)ds} \int_{T_a}^{T_b} h(u)A(T_a,u)e^{-\int_{T_a}^u
\psi(s) ds}\Psi(t,T_a,y_t,y^*,B(T_a,u))du.
\end{equation*}
On the other hand, if
\begin{equation*}
\int_{T_a}^{T_b}\Biggl[ L_{GD} D(T_a,u)
\partial_u \mathbb{S}(T_a,u;0)
+K\mathbb{S}(T_a,u;0)D(T_a,u)\left(1-(u-T_{\beta(u)-1})r_u
\right)\Biggr]du < 0,
\end{equation*}
the default swaption is so deeply in the money that the probability
of it moving out of the money is null. Therefore, in this case the
default swaption is equivalent to a forward default swap, hence it
can be valued by computing the price of the equivalent forward
default swap.

\section{Implementation and numerical results}

To implement the formula presented in the previous section, we need
to compute the relevant integrals numerically. We first focus on
deriving a quadrature formula for computing the integral appearing
in the formula for $\Pi$:
\begin{equation}
\int_0^\infty \frac{e^{U y_0}[S \cos(W y_0+v \varsigma)+R \sin(W
y_0+v \varsigma)]}{v}dv. \label{eq:integral}
\end{equation}

Define $f(v):=S \cos(W y_0+v \varsigma)+R \sin(W y_0+v \varsigma)$.
In the following lemma\footnote{The detailed proof is tedious but straightforward. It is omitted here but is available upon request.}, we confirm that our integrand is continuous and bounded on the
interval $(0,\infty)$ with finite limits on both ends of the
interval.

\begin{lemma}
The function $\frac{e^{U y_0}}{v}f(v)$ is continuous and bounded for
$v \in (0,\infty)$. Moreover,

\begin{eqnarray}
\lim\limits_{v \rightarrow \infty}\frac{e^{U y_0}}{v}f(v) &=& 0,\\
\lim\limits_{v \rightarrow 0}\frac{e^{U y_0}}{v}f(v) &=& C,
\end{eqnarray}
where $C$ is a constant depending on the model parameters.
\end{lemma}

\begin{figure}
    \centering

        \includegraphics[width=14cm,height=6cm]{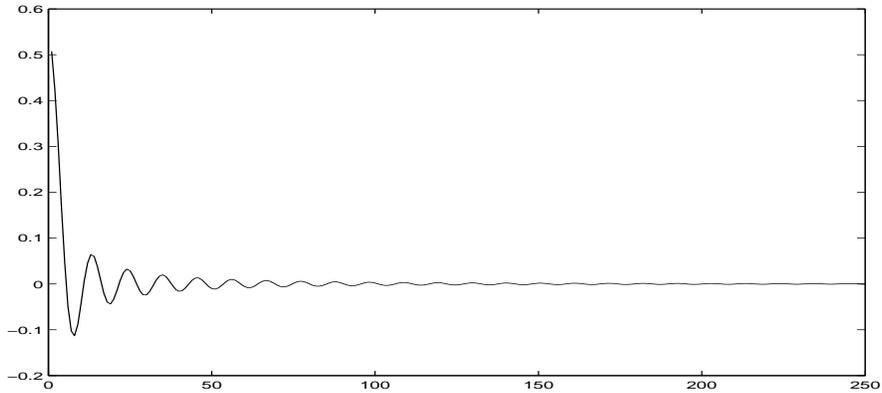}
        \caption{Plot of $\frac{e^{U y_0}}{v} f(v)$ when $v \in (0,250]$ for $y_0=0.005$, $\kappa=0.196$, $\mu=0.065$, $\nu=0.1594$,
$\alpha=0.5$, $\gamma=0.025$, $T-t = 1$} \label{fig:integrand}
\end{figure}

For a visual view of the integrand $\frac{e^{U y_0}}{v}f(v)$, we
plot it for a given set of parameter values in figure
(\ref{fig:integrand}). For the numerical computation of the integral we use the four-point adaptive Gauss-Lobatto quadrature with seven point Kronrod refinement provided by Matlab's "quadl" routine based on Gander and Gautschi (2000). Numerical convergence can be verified in table (\ref{integral}). Experiments -not reported here- against a mid-point trapezoidal and Simpson's quadratures confirmed the accuracy of the faster and more convenient adaptive Gauss-Lobatto algorithm. For the outer integral appearing in the formula (\ref{eq:pso2}), some experimentation shows that Simpson's rule with at worst two quadrature points per quarterly spread payment period is usually enough for convergence of the numerical approximation, while using the spread payment dates as the only quadrature points in most cases leads to a good accuracy.

\begin{table}
\begin{tabular*}{\columnwidth}{@{\extracolsep{\fill}}|l|cccccc|}\cline{1-7}
                     &       &        &         &       &      &        \\
\small{Integral bound: N} & $10^2$   &  $10^3$   &   $10^4$    &   $10^5$     &  $10^6$ & $10^7$
\\
                    &       &        &         &       &      &        \\
\cline{1-7}
                    &       &        &         &       &      &        \\
\small{Numerical integral} & -0.75859  &  -0.76983  &  -0.77173   &  -0.77178   &    -0.77178    &   -0.77178     \\
                    &       &        &         &       &      &        \\
\cline{1-7}
\end{tabular*}							

\caption{Numerical approximation of $\int_{0}^{\infty}\frac{e^{U y_0}}{v} f(v)dv$ by $\int_{0}^{N}\frac{e^{U y_0}}{v} f(v)dv$ using adaptive Gauss-Lobatto quadrature for $y_0=0.005$, $\kappa=0.196$, $\mu=0.065$, $\nu=0.1594$,
$\alpha=0.5$, $\gamma=0.025$, $T-t = 1$, $\varrho = B(0,3)$, $\varsigma=0.0062$}
\label{integral}
\end{table}

\begin{figure}
    \centering

        \includegraphics[width=14cm,height=6cm]{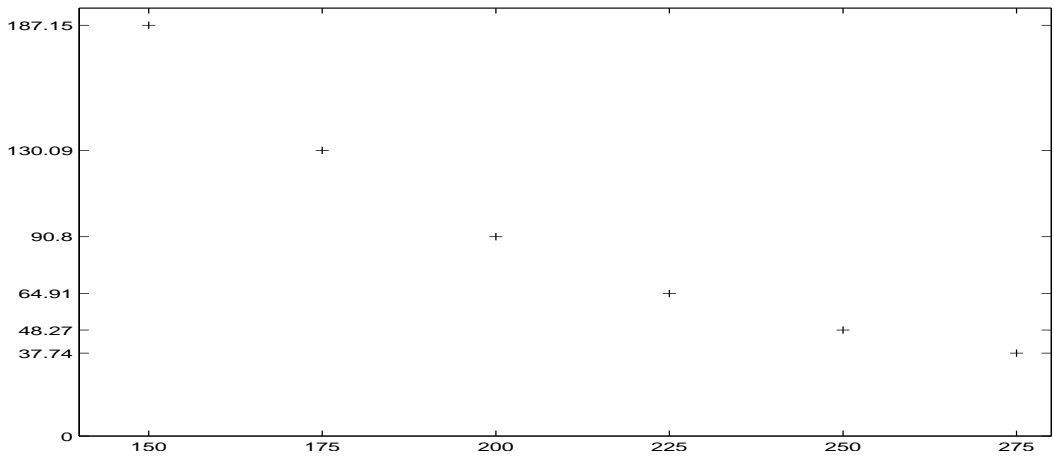}
        \caption{Payer default swaption prices (bps) for different strike values (bps) in the SSRJD model with no deterministic shifts and parameters: $T_a = 1y$, $T_b = 5y$, $r = 0.03$, $L_{GD} = 0.7$, $y_0=0.005$, $\kappa = 0.229$, $\mu = 0.0134$,  $\nu = 0.078$, $\alpha = 1.5$, $\gamma = 0.0067$. The fair value of the underlying forward default swap rate is 204 bps.} \label{fig:optPrices}
\end{figure}

In figure (\ref{fig:optPrices}) we present some numerical results for payer default swaption prices for different strikes, obtained using the quasi-analytic formula developed\footnote{Notice that the jumps in the intensity process can only take positive values. If one thinks in terms of zero-mean shocks, the long term mean reversion level of the process including jumps is no longer the purely diffusive long term mean $\mu$ but the larger $\mu + \frac{\alpha\gamma}{\kappa}$ as summarized in the following equivalent way of writing the dynamics of the process $y_t$: $$dy_t = \kappa(\mu+ \frac{\alpha\gamma}{\kappa}-y_t)dt + \nu\sqrt{y_t}dW_t+(dJ_t-\alpha\gamma dt)$$where the jump increment $dJ_t$ has been centered by subtracting its mean. This is why, in particular, we find a fair value of the underlying forward default swap rate about 204 bps when both the initial condition $y_0$ and the basic long term mean $\mu$ are much smaller. }. These are for a homogenous non-shifted version of the model with constant short rate. The set of parameters used are reported on the figure.



\subsection{Consistency with volatility smile}
In this section, we present some numerical results concerning the
behavior of the model for some parameter values. Our main focus is
on the implied volatility smile that can be generated by the
model. The model potentially allows one to mark-to-market (or
rather mark-to-model) non-ATM default swaptions that may be
present on a trading book. This task cannot be fulfilled with the
market model unless we use ATM implied volatility to value all
options, which should not be acceptable from a risk management
perspective. On the other hand, our intensity model can be
calibrated to the default swap term structure and traded ATM
default swaptions to price other default swaptions more
consistently.

To visualize the implied volatility pattern that can be
generated by the model, we present the numerical results obtained
with three different values of the vector of parameters. The
parameters values are collected in table (\ref{3models_param}).

\begin{table}[htbp]
\begin{tabular*}{\columnwidth}{@{\extracolsep{\fill}}|l|cccccc|}\cline{1-7}
                     &       &        &         &       &      &    \\
\small{Reference} & $y_0$   &  $\kappa$   &   $\mu$    &   $\nu$     &  $\alpha$ & $\gamma$    \\
\cline{1-7}
     &            &               &            &         &         &   \\
Model1   & 0.0007     &    0.4066     &  0.0515    & 0.1507  &  0.5009  &  0.0050  \\
Model2   & 1.3E-06    &    0.4851     &  0.0457    & 0.2000  &  0.5009  &  0.0050  \\
Model3   & 0.005      &    0.2281     &  0.0134    & 0.0782  &  1.5000  &  0.0067   \\
\cline{1-7}
\end{tabular*}
\caption{Three different sets of parameters values of the SSRJD model. }
\label{3models_param}
\end{table}

\begin{figure}[htbp]
\center{\includegraphics[width=14cm,height=6cm]{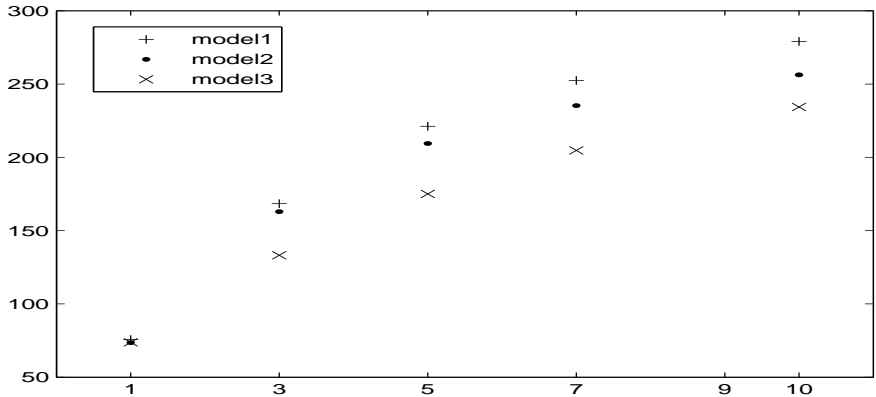}}
\caption{CDS term structures for three sets of parameters values of the SSRJD model. The default swaps have quarterly spread payments. The short rate is assumed to be constant: $r=0.03$. The values of the parameters of the intensity process are given in table (\ref{3models_param}).}
\label{fig:TS_MultModels}       
\end{figure}

We plot in figure (\ref{fig:TS_MultModels}) the CDS term structures
generated by the different models.

\begin{figure}[htbp]
\center{\includegraphics[width=14cm,height=6cm]{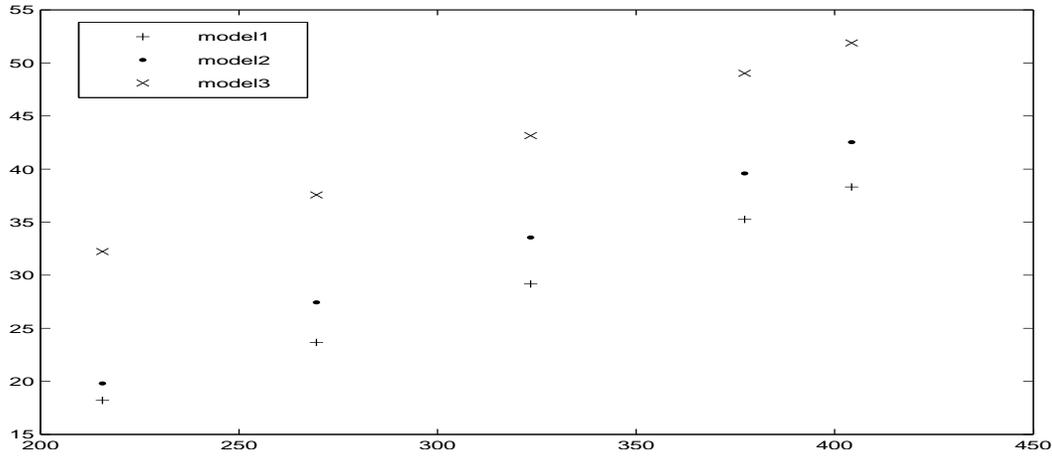}}
\caption{Generated implied volatility smiles for three sets of parameters values of the SSRJD model. The short rate is assumed to be constant: $r=0.03$. The default swaptions have a maturity of one year and can be exercised at maturity into a default swap of a remaining four years to expiry with quarterly spread payments. The values of the parameters of the intensity process are given in table (\ref{3models_param}) and the CDS term structures are plotted in figure (\ref{fig:TS_MultModels}).}
\label{fig:Smile_MultModels}       
\end{figure}

Implied volatility smiles generated from model prices of payer default swaptions with various strikes for
these models are in figure (\ref{fig:Smile_MultModels}). Note that the model implies a plausible upward sloping volatility smile. It is also noticeable that the presence of a significant jump component can result in higher implied volatilities than a model with a less important jump component even when the last one results in a steeper CDS term structure.

\section{Concluding remarks}
The SSRJD model can fit the current default swap term structure while being consistent with some dynamic future deformations and implying a volatility smile for default swaptions. The quasi-analytic formula presented in this paper permits fast and accurate pricing of default swaptions. Hence, the model could be calibrated to the CDS term structure and a few default swaptions, to price and hedge other credit derivatives consistently.

\begingroup
     \parindent 0pt
     \parskip 2ex
     \def\enotesize{\normalsize}
    \theendnotes
   \endgroup

\end{document}